\documentstyle[12pt,aaspp4,epsfig]{article}

\def\ls{{_<\atop^{\sim}}}
\def\gs{{_>\atop^{\sim}}}
\def\cgs{{ erg cm$^{-2}$ s$^{-1}$}}

\begin{document}

\title{The BeppoSAX view of the X-ray active nucleus of NGC4258}

\author{F. Fiore$^{1,2,3}$, S. Pellegrini$^4$, G. Matt$^5$,
L.A.~Antonelli$^1$, A.~Comastri$^6$, R.~Della Ceca$^7$, 
E.~Giallongo$^1$, S.~Mathur$^8$, S.~Molendi$^9$, A.~Siemiginowska$^3$, 
G.~Trinchieri$^7$,\\ and B.~Wilkes$^3$\\
$^1$ Osservatorio Astronomico di Roma, Via Frascati 33,
I-00044 Monteporzio, Italy\\
$^2$ BeppoSAX Science Data Center, Via Corcolle 19, I-00131 Roma, Italy\\
$^3$ Harvard-Smithsonian Center of Astrophysics, 60 Garden Street, 
Cambridge MA 02138 USA\\
$^4$ Dip. di Astronomia, Universit\`a di Bologna, via Ranzani 1, 
I-40127, Bologna,  Italy\\
$^5$ Dip. di Fisica, Universit\`a Roma Tre, via della Vasca 
Navale 84, I-00146, Roma, Italy\\
$^6$ Osservatorio Astronomico di Bologna, via Ranzani 1, I-40127, Bologna, 
Italy\\
$^7$ Osservatorio Astronomico di Brera, Via Brera 28, 20121, Milano, Italy\\
$^8$ Astronomony Dept. The Ohio State University, 140 West 18th Avenue
Columbus, OH 43210, USA \\
$^9$ IFC/CNR, via Bassini 15, Milano, Italy
}

\author{\tt (version: 12 February 2001) }

\begin{abstract}
BeppoSAX observed the Seyfert 1.9 galaxy NGC4258 on December 1998,
when its 2-10 keV luminosity was about $10^{41}$ erg s$^{-1}$.  Large
amplitude ($100\%$) variability is observed in the 3-10 keV band on
timescales of a few tens of thousands seconds while variability of
$\sim20\%$ is observed on timescales as short as one hour.  The
nuclear component is visible above 2 keV only, being obscured by a
column density of $(9.5\pm1.2)\times10^{22}$ cm$^{-2}$; this component
is detected up to 70 keV with a signal to noise $\gs3$, and with the
steep power law energy spectral index of $\alpha_E=1.11\pm0.14$.
Bremsstrahlung emission for the 2-70 keV X-ray luminosity, as expected
in Advection Dominated Accretion Flow (ADAF) models with strong winds,
is ruled out by the data.  The ratio between the nuclear radio (22
GHz) luminosity and the X-ray (5 keV) luminosity is $\ls10^{-5}$,
similar to that of radio-quiet quasars and Seyfert galaxies.  X-ray
variability and spectral shape, radio-to-X-ray and near
infrared-to-X-ray luminosity ratios suggest that the nucleus of
NGC4258 could be a scaled-down version of a Seyfert nucleus, and that
the X-ray nuclear luminosity can be explained in terms of
Comptonization in a hot corona.

The soft ($E\ls2$keV) X-ray emission is complex. There are at least
two thermal-like components with temperatures of $0.6\pm0.1$ keV and
$\gs1.3$ keV. The cooler ($L_{0.1-2.4keV}\sim10^{40}$ erg s$^{-1}$)
component is probably associated with the jet, resolved in X-rays by
the ROSAT HRI (Cecil et al. 1994).  The luminosity of the second
component, which can be modeled equally well by an unobscured power
law model with $\alpha_E=0.2^{+0.8}_{-0.2}$, is
$L_{0.1-2.4keV}\sim7\times10^{39}$ erg s$^{-1}$, consistent with that
expected from discrete X-ray sources (binaries and SN remnants) in the
host galaxy.

Observations of NGC4258 and other maser AGNs show strong nuclear X-ray
absorption. We propose that this large column of gas might be
responsible for shielding the regions of water maser emission from
X-ray illumination.  So a large column density absorbing gas may be a
necessary property of masing AGNs.

\end{abstract}

 
\section{Introduction}

The nearby (distance of 7.2$\pm$0.3 Mpc, Herrnstein et al. 1999),
bright ($B_0^T=8.5$), SABbc galaxy NGC4258 (M106) is spectroscopically
classified as a 1.9 Seyfert galaxy (Ho, Filippenko \& Sargent 1997,
who find [OIII]/H$\beta$=10.3, and [SII]/H$\alpha$=0.94).  Wilkes et
al. (1995) find relatively broad (1000 km s$^{-1}$) emission lines,
which are strongly polarized (5-10\%), supporting the existence of an
obscured active nucleus in this galaxy.  Further support to this case
comes from high resolution observations of the water maser in the
galaxy nucleus.  Miyoshi et al. (1995) discovered a highly inclined
thin disk between 0.13 and 0.25 pc. The inclination of the disk is
estimated to 82$\pm$1 degrees (Herrnstein et al. 1999, the inclination
of the galaxy is 70 degrees, Ho et al. 1997). The aspect ratio of the
disk is extremely small, $\ls0.2\%$ (Herrnstein et al. 1999).  The
Keplerian rotation curve traced by the water maser requires a central
binding mass, presumably in the form of a supermassive black hole, of
$3.9(\pm0.1)\times10^7$ M$_{\odot}$ (Herrnstein et al. 1999).  To
date, this is the strongest evidence for a supermassive black hole in
a galaxy and the best measurement of its mass. The 
implied Eddington luminosity is $\sim5\times10^{45}$ erg s$^{-1}$. 
Pending an accurate estimate of the observed nuclear luminosity, it
is therefore possible in this case to constrain tightly the $L/L_{Edd}$
ratio of the active nucleus.

The evaluation of the nuclear luminosity and spectrum is not an easy
task because the nucleus is highly obscured.  Wilkes et al. (1995)
discovered a faint, blue, highly polarized nuclear continuum, but the
uncertainty on the optical nuclear luminosity is large (two orders of
magnitude, depending on the nature of the scattering region). Similar
uncertainties are found using the intensity of the emission
lines. Chary \& Becklin (1997) found a ``nuclear'' emission at 2
micron, in excess of that expected from the galaxy profile, of 4.5
mJy, corresponding to a luminosity ($\nu L(\nu)$) of about $10^{41}$
erg s$^{-1}$.  The nucleus is not detected in VLBA radio
maps. Herrnstein et al. (1998) report upper limits to the 22 GHz flux
and luminosity of 0.22 mJy and $\nu L(\nu)=3\times10^{35}$ erg
s$^{-1}$ respectively. Conversely, a double, twisted radio jet is
visible on scales from milliarcsec to a few arcmin (the latter scale
corresponding to a few kiloparsec at the galaxy distance).  The jets are
visible in optical and X-ray images too. In particular, Cecil, Wilson
\& De Pree (1995) resolved the jets in ROSAT HRI observations and
measured a 0.1-2.4 keV luminosity of $6.5\times10^{39}$ erg s$^{-1}$.
They also detected an unresolved ``nuclear'' emission of comparable
luminosity, confirming previous Einstein results (Fabbiano et
al. 1992).  ASCA spectra (Makishima et al. 1994, Ptak et al. 1999,
Reynolds et al. 2000) resolve at least two distinct components, a
thermal component dominating the spectrum up to 2-3 keV and a highly
absorbed power law dominating the spectrum above this energy.  The
ASCA power law energy index is $\alpha_E\sim0.8$ and the neutral
absorbing column is in the range $N_H=(0.9-1.5)\times10^{23}$
cm$^{-2}$ (see Reynolds et al. 2000 for details).  The 2-10 keV flux
observed by ASCA varied between $3\times10^{-12}$ and $10^{-11}$ \cgs
(Reynolds et al. 2000), (6-15)$\times10^{-12}$ \cgs, when corrected
for intrinsic absorption, during several observations performed
between May 1993 and May 1999. This implies a 2-10 nuclear luminosity
in the range $(0.4-1)\times10^{41}$ erg s$^{-1}$.  The uncertainties on
the spectral index and absorbing column are quite large (0.2--0.3 and
$(1-2)\times10^{22}$ cm$^{-2}$ respectively), due to the fact that
only a narrow energy band is unaffected by the bright thermal
component, and so is the uncertainty on the unobscured flux. Reynolds
et al. (2000) also report the detection of a narrow 6.4 keV iron
K$\alpha$ line of equivalent width of $107^{+42}_{-37}$ eV.
 
Because of the small luminosity in terms of the Eddington one, the
nuclear emission of NGC4258 can be explained either by a standard
accretion disk with a very small accretion rate, or by radiatively
inefficient accretion models, such as the Advection Dominated
Accretion Flows (ADAF, e.g. Narayan \& Yi 1995).  The Radio to X-ray
Spectral Energy Distribution (SED) has been interpreted in terms of
emission from an ADAF with a critical accretion rate $\dot m \sim
0.016\alpha_{-1}$ (Lasota et al, 1996; $\alpha_{-1}$ is the viscosity
parameter in units of 0.1) or $\dot m \sim0.012\alpha_{-1}$ (Gammie et
al. 1999, their modeling also includes the NIR data and the 22 GHz
upper limit).  In ADAF models the radio luminosity comes from
synchrotron emission and the X-ray comes from the combined
contributions of a high temperature (kT$\gs$100 keV) thermal
bremsstrahlung and Comptonization of the synchrotron photons.  Lasota
et al. (1996) point out that this representation is not unique,
because of the substantial uncertainties in the optical-UV and X-ray
data and because of the difficulty in distinguishing an ADAF from
thermal Comptonization, using only the X-ray spectral shape.
Both Lasota et al. (1996) and Gammie et al. (1999) do not account in
their models for the strong wind expected from the inner accretion
disk (Blandford \& Begelman 1999, Di Matteo et al. 2000).  The wind
carries a significant fraction of the mass, energy and angular
momentum. In particular, it drastically reduces the synchrotron
emission in the radio band and the Compton emission in the O-UV and
X-ray bands (Di Matteo et al.  2000).  In both modelings the NIR
luminosity is generated by a standard, geometrically thin disk,
external to the ADAF transition radius.

BeppoSAX, with its good sensitivity over a broad (0.1-200 keV) band,
can help in separating the different X-ray spectral components and can
provide a strong constraint on the nuclear power law spectral index
and on the high energy cut-off (predicted by both ADAF models and
thermal Comptonization models). It can help understanding whether the
NGC4258 nucleus hosts a ``normal'' AGN (although of low luminsity) or
rather an ADAF. BeppoSAX observed NGC4258 for about 100 ks on December
19-22 1998.  The nuclear component was in a high state in comparison
with most ASCA observations. This, together with the broad band (the
source is detected with a signal to noise ratio $\gs 3$ up to about 70
keV) allows us to tightly constrain the flux and spectral shape of the
nuclear hard component.

The paper is organized as follows: Section 2 presents the data and
gives information on their reduction; Section 3 presents a variability
analysis; Section 4 presents the spectral analysis and the broad band,
radio to X-ray spectral indices; Section 5 discusses the main
results on the hard nuclear component and on the low energy
components.

\section{Observation and data reduction}

The observations were performed with the BeppoSAX Narrow Field
Instruments, LECS (0.1-10 keV, Parmar et al. 1997), MECS (1.3-10 keV,
Boella et al. 1997b), HPGSPC (4-60 keV, Manzo et al. 1997) and PDS
(13-200 keV, Frontera et al. 1997).  LECS and MECS are imaging gas
scintillation proportional counters, the HPGSPC is a collimated high
pressure gas scintillation proportional counter and the PDS consists
of four phoswich units. The PDS is operated in the so called ``rocking
mode'', with a pair of units pointing to the source while the other
pair monitor the background $\pm210$ arcmin away. The units on and off
source are interchanged every 96 seconds. We report here the analysis
of the LECS, MECS and PDS data; the HPGSPC data are very noisy, due to
high HPGSPC background.  The MECS observations were performed with
units 2 and 3 (on 1997 May 6$^{\it th}$ a technical failure caused the
switch off of unit MECS1); these data were combined together after
gain equalization. The LECS is operated during dark time only,
therefore LECS exposure times are usually smaller than MECS ones (a
factor of 3 for NGC4258).  Table 1 gives the LECS, MECS and PDS
exposure times and the count rates.

Standard data reduction was performed using the SAXDAS software
package version 2.0 following Fiore, Guainazzi and Grandi (1999).  In
particular, data are linearized and cleaned from Earth occultation
periods (we accumulated data for Earth elevation angles $>5$ degrees)
and unwanted periods of high particle background (satellite passages
through the South Atlantic Anomaly and periods with magnetic cut-off
rigidity $>6$ GeV/c).  The internal background of the LECS, MECS and
PDS during the accepted periods is relatively low and stable
(variations at most of 30\% during the orbit) due to the low
inclination of the satellite orbit (3.95 degrees).  LECS and MECS
spectra were extracted from regions of 8 arcmin and 3 arcmin radii
respectively. These radii maximize the signal-to-noise ratio below 1
keV in the LECS and above 2 keV in the MECS. Background spectra were
extracted in detector coordinates from high Galactic latitude `blank'
fields (98\_11 issue) using regions equal in size to the source
extraction region. We have checked that the mean level of the
background in the LECS and MECS ``blank fields'' observations is
comparable with the mean level of the background in the NGC4258
observations using source free regions at various positions in the
detectors.

The PDS data have been reduced using the variable Risetime threshold
technique to reject particle background (see Fiore, Guainazzi \&
Grandi 1999). This technique reduces the total 13-200 keV background
to about 20 counts s$^{-1}$ and the 13-80 keV background to about 6
counts s$^{-1}$ (instead of 30 and 10 counts s$^{-1}$ respectively,
obtained using the standard fixed Risetime threshold technique).  The
PDS rocking mode provides a very reliable background subtraction.
This can be checked looking at the spectrum between 200 and 300 keV,
where the effective area of the PDS to X-ray photons is small and
therefore the source contribution is negligible; we obtain after
background subtraction 0.0085$\pm$0.012 counts s$^{-1}$, consistent
with the expected value of 0.  The net (background subtracted) 13-130
keV on-source signal is 0.121$\pm$0.024 counts s$^{-1}$.  The source
is detected with a signal to noise ratio $\gs3$ up to $\sim 70$
keV. The 13-130 keV PDS count rate is $\gs3.5$ times the systematic
uncertainty in the PDS background subtraction, equal to
0.020$\pm$0.015, (Guainazzi \& Matteuzzi 1997).  Confusion in the PDS
collimator Field Of View (1.4 degrees FWHM) ultimately limits our
capability to constrain the high energy spectrum. We have carefully
checked for any possible contaminant in a region of 1.5 degrees radius
around NGC4258 (using the NED, SIMBAD, AGN, clusters, CVs, Radio and
X-ray sources catalogs) finding no obvious bright hard X-ray source.
Of course there is the possibility of a bright ``unknown'' source in
the PDS field of view.  The chance of finding a source in any given
2~square degrees, the PDS beam area, is however small.  The HEAO-1 A4
all sky catalog (Levine et al. 1984) lists just 7 high Galactic
latitude sources in the 13-80~keV band down to a flux of
2$\times$10$^{-10}$
\cgs (10mCrab). The 13-80 keV flux is about 20 times smaller than this
figure and so, assuming a logN-logS slope of $-$1.5, we expect a
chance coincidence rate of 2\%.

\begin{table}[ht]
\caption{\bf Observation log for BeppoSAX sequence 50491001}
\begin{tabular}{lccc}
\hline
\hline
Instrument & Exposure (ks) & Band & Count rate \\
\hline
LECS & 33.3 & 0.1-4 keV  & 3.00$\pm$0.11$^a$  \\
MECS & 99.4 & 1.7-10 keV & 7.25$\pm$0.09$^a$  \\
PDS  & 46.9 & 13-130 keV & 0.121$\pm$0.024 \\
\hline
\end{tabular}

$^a$ $10^{-2}$ counts s$^{-1}$
\end{table}

\section{Variability}

\begin{figure*}
\centerline{ 
\psfig{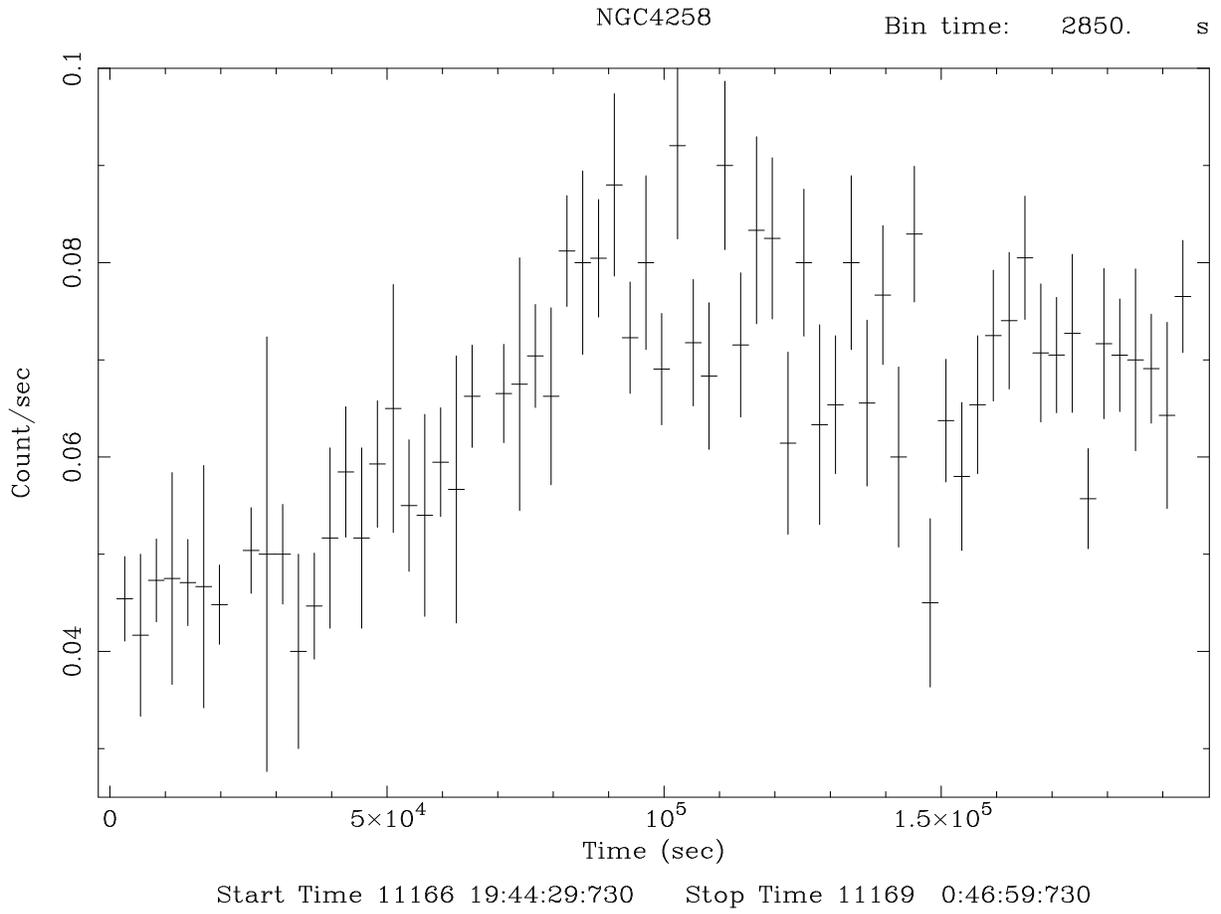}
}
\caption{ 
The BeppoSAX MECS 3-10 keV light curves of NGC4258 in bins
of 2850 seconds (half satellite orbit). Errors represent the
1 $\sigma$ confidence interval.
}
\label{totlc}
\end{figure*}

Figure \ref{totlc} shows the MECS 3-10 keV light curve in bins of 2850
seconds.  Variability of a factor of about two is evident on
timescales of half day, as well as smaller amplitude variations
(10-20~\%) on time scales as short as one hour. Shorter timescales
variations of similar amplitude cannot be sampled, due to the limited
sensitivity of the MECS instrument.  In this regard the source behaves
in a way similar to X-ray brighter, and better studied, Seyfert
galaxies (e.g. NGC4051, NGC5506, Green et al. 1993, Nandra et
al. 1997).

\begin{figure*}
\centerline{ 
\psfig{figure=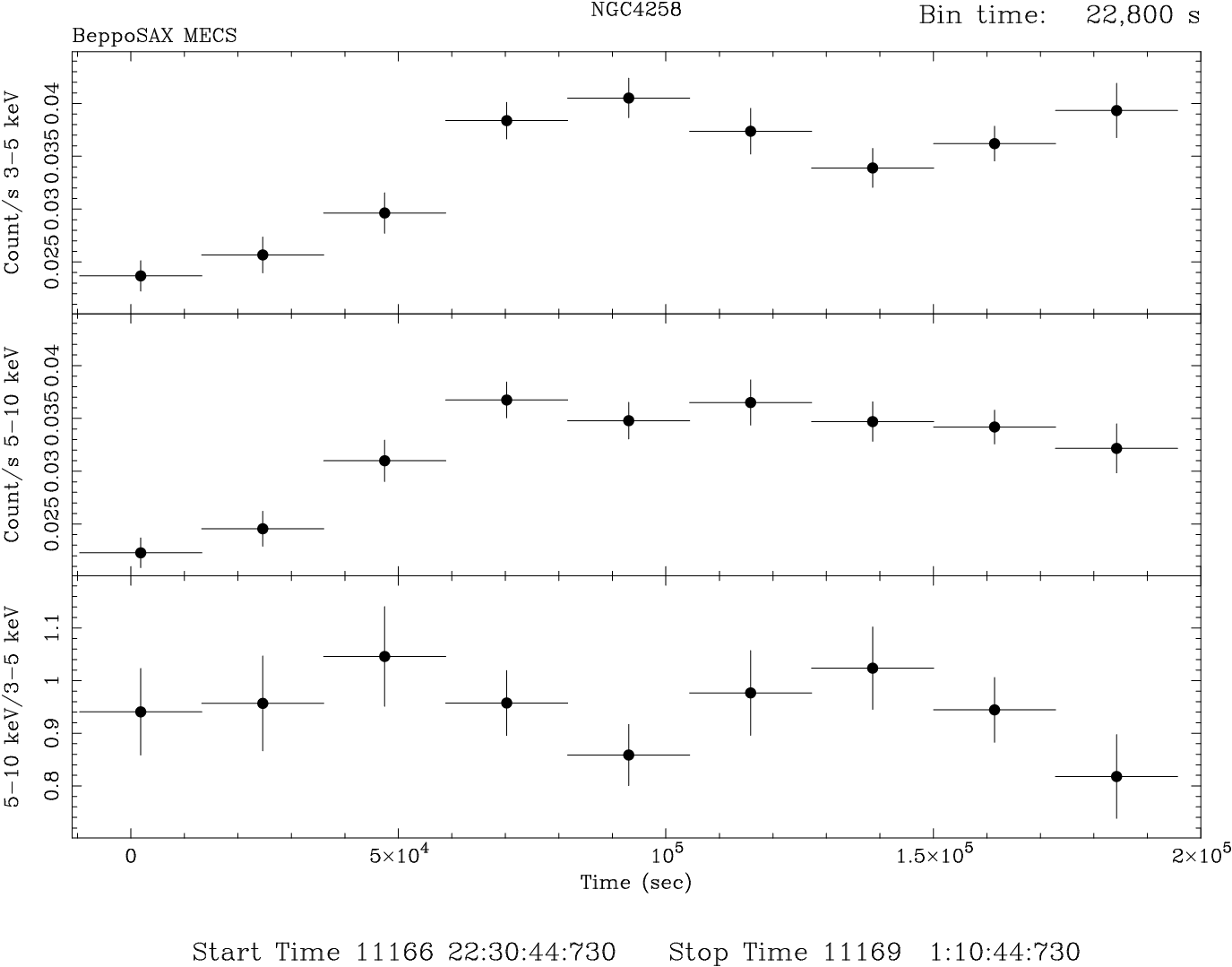,height=14cm,width=16cm,angle=-90}
}
\caption{ 
The BeppoSAX MECS light curves of NGC4258 in the 3-5 keV and
5-10 keV bands (upper and middle panels, bins of 22800 seconds, 
corresponding to about 4 satellite orbits per bin). The 5-10keV/3-5keV
count ratio as a function of time (lower panel).
}
\label{melc}
\end{figure*}

Figure \ref{melc} shows the MECS 3-5 keV and 5-10 keV light curves (on
the same scale), together with their ratios.  The light curves are not
background subtracted (the internal+cosmic background is about 3\% and
6\% of the mean total count rate in the two bands, and therefore it is
negligible).  The two MECS lightcurves present similar but not
identical variations and, in fact, their hardness ratio shows small
($\ls20\%$) variations.  Unfortunately the statistic is not good
enough to perform an accurate spectral variability study.  To this
purpose high throughput instruments, like those that are on board the
Chandra and XMM-Newton satellites, are needed.  In any case, the
spectral variations are expected to be small enough to allow us to
safely use time integrated counts.  We therefore concentrate
hereinafter on the time averaged properties of the spectrum.

\section {Spectral analysis}

Spectral fits were performed using the XSPEC 9.0 software package and
public response matrices as from the 1998 November release.  LECS and
MECS spectra were rebinned following two criteria: a) to sample the
energy resolution of the detectors with four channels at all energies
whenever possible, and b) to obtain at least 20 counts per energy
channel.  Constant factors have been introduced in the fitting models
in order to take into account the intercalibration systematics between
the instruments. The expected factor between LECS and MECS is about
0.9 [0.7-1.1]; the factor between the PDS and MECS is 0.8 [0.7-0.9]
(see Fiore, Guainazzi \& Grandi 1999). We assume in all spectral fits
the MECS as reference instrument. The energy range used for the fits
are: 0.1-4 keV for the LECS (channels 11-400), 1.65-10 keV for the
MECS (channels 37-213) and 13-100 keV for the PDS. Errors quoted in
this paper are 90\% confidence intervals for one interesting parameter
($\Delta\chi^2=2.71$).

\begin{figure*}
\centerline{ 
\psfig{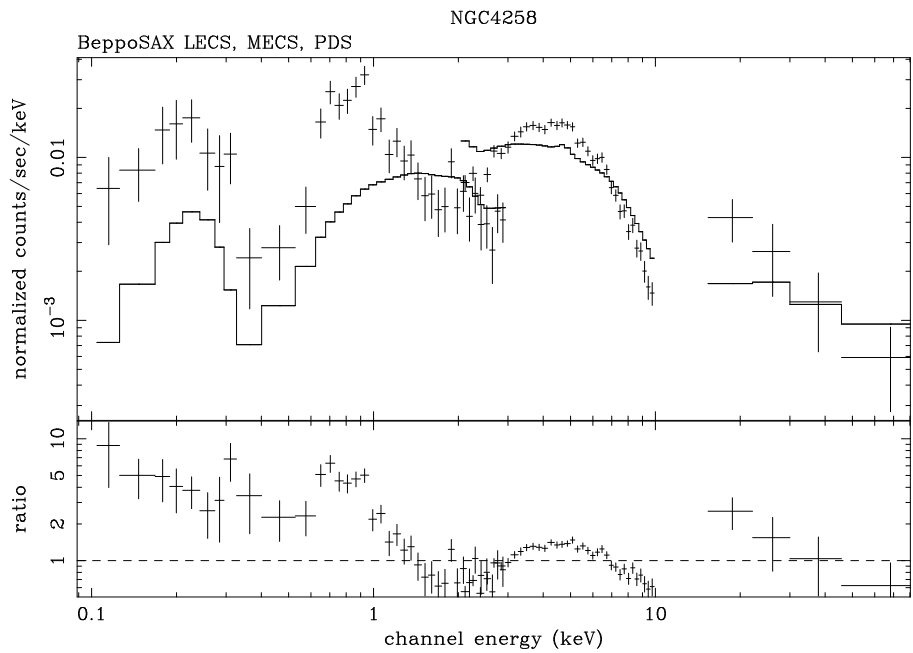}
}
\caption{ 
The 0.1-100 keV BeppoSAX LECS, MECS and PDS spectra of NGC4258
fitted with a single power law model in the full 0.1-100 keV band. 
Note the strong cut-off below 4 keV and the large
excess between 0.7 and 1 keV (likely due to iron L emission
from an optically thin plasma).
}
\label{spe1}
\end{figure*}

Figure \ref{spe1} shows that the 0.1-100 keV spectrum (LECS+MECS+PDS)
is highly complex. A strong cut-off below 4 keV is clearly visible as
well as a faint emission line feature between 6 and 7 keV.  The large
excess visible between 0.7 and 1 keV is likely due to iron L emission
from an optically thin plasma.  The next Section presents the analysis
of the high energy spectrum (E$>2.5$ keV). The thermal component(s)
dominating the spectrum below $\sim2$ keV are discussed in Section
4.2.

\subsection {The hard nuclear component}

\begin{figure*}
\centerline{ 
\psfig{figure=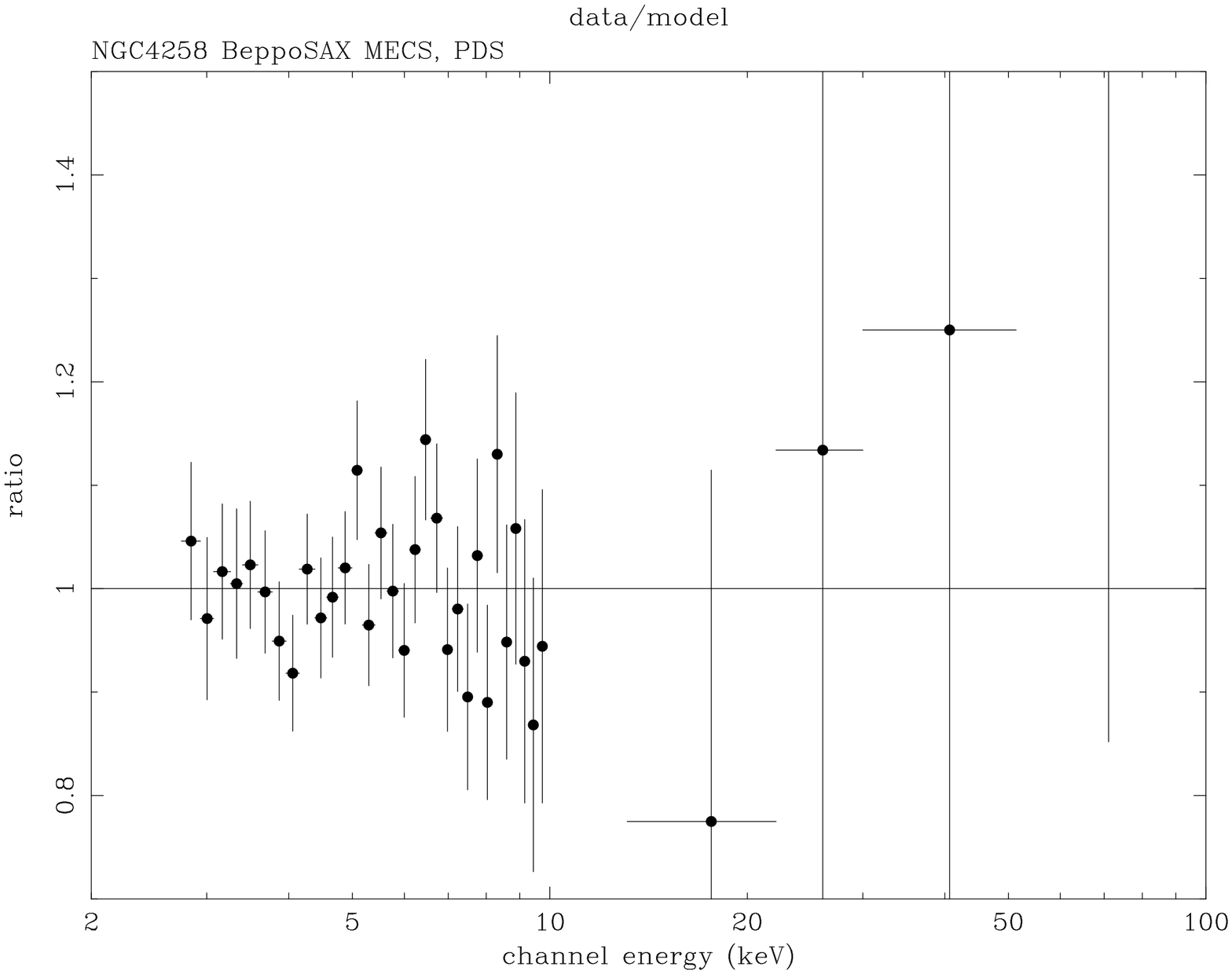,height=12cm,width=8.5cm,angle=-90}
}
\caption{ 
The ratio between the MECS and PDS 2.5-100 keV data and the
best fitting absorbed power law model given in Table 2.
}
\label{ratio1}
\end{figure*}

The intensity of the hard component is strongly reduced by
photoelectric absorption below 2-3 keV. For the sake of simplicity we
therefore exclude the LECS data and limit the fit to the MECS+PDS data
above 2.5 keV, when considering the nuclear component.  The MECS+PDS
spectra were fitted with a power law model plus photoelectric
absorption. The agreement between the data and this simple model is
acceptable (see Table 2).  Thanks to the broad energy band the power
law spectral index and the column density of the absorbing gas are
well constrained ($\alpha_E=1.11\pm0.14$, $N_H=0.95(\pm0.12)
\times10^{23}$ cm$^{-2}$).  Figure \ref{ratio1} shows the ratio
between the MECS and PDS data and the best fit model.  

The observed 2-10 keV flux is $8.0\times10^{-12}$ \cgs, while the flux
corrected for intrinsic absorption is $1.52(\pm0.15)\times10^{-11}$
\cgs.  The uncertainty on the unabsorbed flux is mostly due to the
uncertainty on the absorbing column. The 2-10 keV nuclear (unobscured)
luminosity is $1.0(\pm0.1)\times10^{41}$ erg s$^{-1}$.  The broad band
0.1-100 keV luminosity of the hard component is $4.5\times10^{41}$ erg
s$^{-1}$.

The inclusion of a narrow line is significant at the 98\% confidence
level according to the F test.  The line energy (6.57$\pm$0.20 keV) is
consistent with K$\alpha$ emission from both neutral and helium like
iron. The equivalent width (85$\pm$65 eV) is not well constrained. It
is consistent with but somewhat lower than that of Seyfert 1 and
Compton thin Seyfert 1.9-2 galaxies (Matt 2000).  Assuming neutral
iron, the observed equivalent width is consistent with that expected
from transmission through a column density of about $10^{23}$
cm$^{-2}$ (Ghisellini, Haardt \& Matt 1994).
The inclusion of a high energy cut-off is not significant.  The 90\%
confidence level lower limit to the cut-off energy is 30 keV.

We have also fitted the MECS and PDS data with a thermal
bremsstrahlung model (table 2). The $\chi^2$ is significantly worse
than in the power law model fit. We have then fitted the data with a
power law plus bremsstrahlung model (fixing the temperature of the
bremsstrahlung to 100 keV), to put a limit on the flux of an
additional spectral component with the shape predicted by ADAF models
with strong winds.  The 90\% limits on the 2-10 keV unabsorbed flux
and on the 0.1-100 keV luminosity of this component are
$4\times10^{-12}$ \cgs and $1.5\times10^{41}$ erg s$^{-1}$
respectively, about one third of the total 0.1-100 keV nuclear
luminosity.

\begin{table*}[ht]
\caption{\bf Spectral fits: 2.5-100 keV band}
\footnotesize
\begin{tabular}{lcccccc}

\hline
\hline
model & $N_H^a$ & $\alpha_E$ or kT$^b$ & line E$^b$ & 
line EW$^c$ & $\chi^2$(dof) \\ 
\hline
Power Law & 9.5$\pm$1.2 & 1.08$\pm$0.14 &  & & 30.3(40)  \\
Power Law + line & 9.4$\pm$1.2 & 1.11$\pm$0.14 & 6.57$\pm$0.20 & 
85$\pm$65 & 25.3(38) \\
Brem.     & 7.9$^{+0.7}_{-1.3}$ & 9.5$^{+3.5}_{-1.2}$ & & & 40.7(44)\\ 
\hline
\end{tabular}

\normalsize

$^a$ in $10^{22}$ cm$^{-2}$; $^b$ in keV; $^c$ in eV
\end{table*}

\begin{table*}[ht]
\caption{\bf Spectral fits: 0.1-100 keV band}
\footnotesize
\begin{tabular}{lccccccc}
\hline
\hline
model &  $kT_1^a$ & $kT_2^a$ or $\alpha_{E2}$ & A$^b_1$ & A$^b_2$ & 
Flux$_1^c$ & Flux$_2^c$ & $\chi^2$(dof) \\ 
\hline
MEKAL+PL+line & 0.81$\pm$0.07 & -- & 0.11$\pm0.05$ & &  2.0 & -- & 
108.9 (92) \\
2 MEKAL+PL+line & $0.6^{+0.07}_{-0.20}$ & $>1.3$ & $>0.2$ 
& $<0.9$ & 1.7, 0.03 & 1.1, 0.6 & 87.3(89) \\
MEKAL+2PL+line & 0.6$\pm0.1$ & $0.2^{+0.8}_{-0.2}$ & $>0.2$ & -- & 1.6, 
0.04 & 0.4, 2.0 & 85.3 (90)\\
MEKAL+2PL+line$^d$ & 0.6$\pm0.1$ & 1.05$\pm0.08$ & $>0.2$ & -- & 1.0, 
0.02 & 1.3, 0.90 & 89.6 (91)\\
\hline

\end{tabular}

\normalsize

$^a$  in keV; 
$^b$ metal abundance in unit of solar abundance;
$^c$ 0.1-2.4 and 2-10 keV fluxes in $10^{-12}$ \cgs;
$^d$ the spectral indices of the obscured and unobscured power laws are
linked together

\end{table*}

\subsection {The low energy spectrum}

To study the low energy spectrum of NGC4258 we now include in the
spectral fitting the MECS data in the full 1.7-10 keV band and the
LECS spectrum, which covers the 0.1-4 keV band.  The LECS, MECS and
PDS data were fitted with an optically thin thermal plasma model
(MEKAL) plus the nuclear absorbed power law component and an iron
K$\alpha$ line. An additional absorbing column of density fixed to the
Galactic value along the line of sight ($N_H=1.16\times10^{20}$
cm$^{-2}$) has been included in the fit.  The best fit power law
spectral index and intrinsic absorption resulted very close to the
best fit values found in the previous section.  In Table 3 we give
only the parameters of the low energy component.

\begin{figure*}
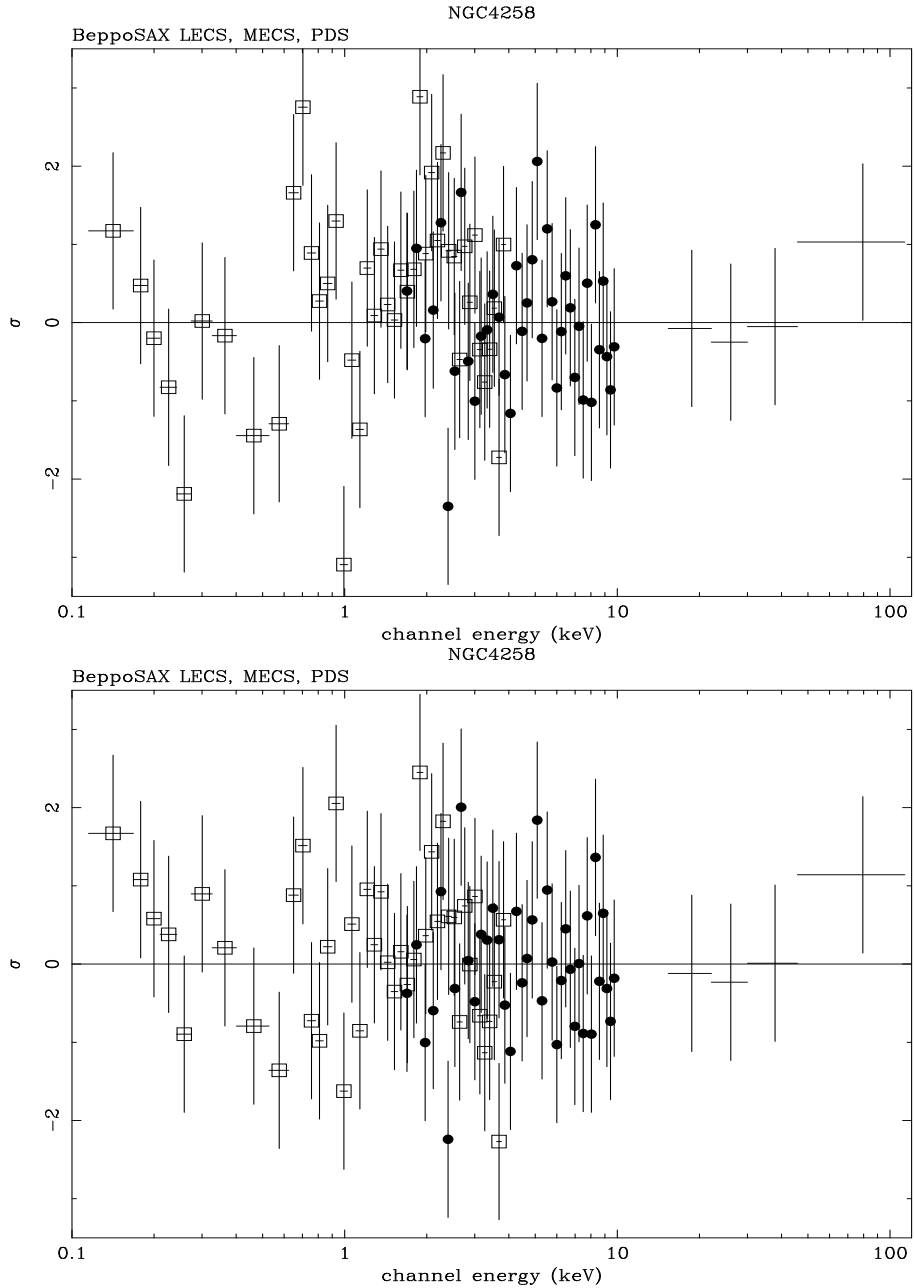

\centerline{ 
\vbox{
\psfig{figure=f5a.eps,height=12cm,width=8.5cm,angle=-90}
\psfig{figure=f5b.eps,height=12cm,width=8.5cm,angle=-90}
}
}
\caption{ Upper panel,
the residuals, in number of $\sigma$, after the subtration of the 
best fit single MEKAL plus power law 
model from the LECS (open squares), MECS (filled circles) and
PDS (crosses) data. Note the relatively large deviations around 0.7 and
1 keV; lower panel, same as in the upper panel but for a model including
two thermal components. The residuals do not show any large deviation 
in the whole 0.1-100 keV band.
}
\label{lecs_spe1}
\end{figure*}

The fit with a single thermal component is not completely satisfactory
(see Figure \ref{lecs_spe1}, upper panel), with rather large residuals
between 0.5 and 1 keV.  The LECS and MECS spectra have therefore been
fitted with a two temperature model plus the highly absorbed power law
and the iron K$\alpha$ line (figure \ref{lecs_spe1}, lower panel).
The improvement in $\chi^2$ is significant ($\Delta\chi^2=21.6$ for
three additional parameters, corresponding to a probability of 99.98\%, 
according to the F test, see Table 3).  While the temperature of
the cooler component is well constrained, for the hotter component we
obtain only a lower limit of about 1.3 keV.  The metal abundance of
the cooler component is constrained to be higher than 0.2 solar, while
that of the hot component is constrained to be smaller than 0.9 solar.
The results on the other parameters do not change fixing the metal
abundances to the solar value.  The 0.1-2.4 keV luminosities of the
cool and hot components are about 10 and 7 $\times10^{39}$ erg
s$^{-1}$, respectively.  The 2-10 keV luminosity of the hot component
is $\sim4\times10^{39}$ erg s$^{-1}$.  Because of the complexity of
the model and of the limited energy resolution of the LECS and the
MECS the uncertainties on these values are however rather large, the
order of 50\%.  A slightly better fit is obtained by replacing the
thermal ``hot'' component with an unobscured power law.  In this case
the improvement in $\chi^2$ is significant at the 99.998\% confidence
level ($\Delta\chi^2=23.6$ for two additional parameters).  Fixing the
spectral index of the unobscured power law to that of the obscured
nuclear component (so mimicking a partial covering model) produces a
$\chi^2$ of 89.6, still acceptable. The covering fraction is in this
case $0.935\pm0.010$.

\subsection{Comparison with previous ROSAT and ASCA observations}

The BeppoSAX 2-10 keV flux is about 2 times higher than in the May
1993 and May 1999 ASCA observations, while it is similar to that in
the 1996 observations (Reynolds et al. 2000). The power law index is
consistent with the ASCA measurements.  The column density measured by
Makishima et al. (1994) is $1.5(\pm0.2) \times10^{23}$ cm$^{-2}$,
while Ptak et al. (1999) find a nominally lower value
($0.5-0.7)\times10^{23}$ cm$^{-2}$ with a typical statistical error of
$(+0.8-0.3)\times10^{23}$ cm$^{-2}$, see their tables 5 and 7.
Reynolds et al. (2000) report values between 0.88 and
$1.4\times10^{23}$ cm$^{-2}$. The BeppoSAX measurement is consistent
with all ASCA measurements except that in the May 1993 observation.

The temperatures and abundances of the low energy components are
consistent with ASCA (Makishima et al. 1994, Ptak et al. 1999,
Reynolds et al. 2000) and ROSAT (Pietsch et al. 1994, Cecil, Wilson
and De Pree 1995) determinations; the luminosity of the ``cool''
component is in agreement with the PSPC, HRI and ASCA measurements.
The luminosity of the ``cool'' and ``hot'' components are slightly
higher than in the ROSAT PSPC and HRI observations and in the ASCA
observations, if taken at the face value, but the uncertainties in
both the BeppoSAX data on these luminosities are rather large (30 to
50\%).  The energy and equivalent width of the iron line are
consistent with the ASCA determination.

\subsection {The radio to X-ray nuclear Spectral Energy Distribution}

\begin{figure}
\centerline{ 
\psfig{figure=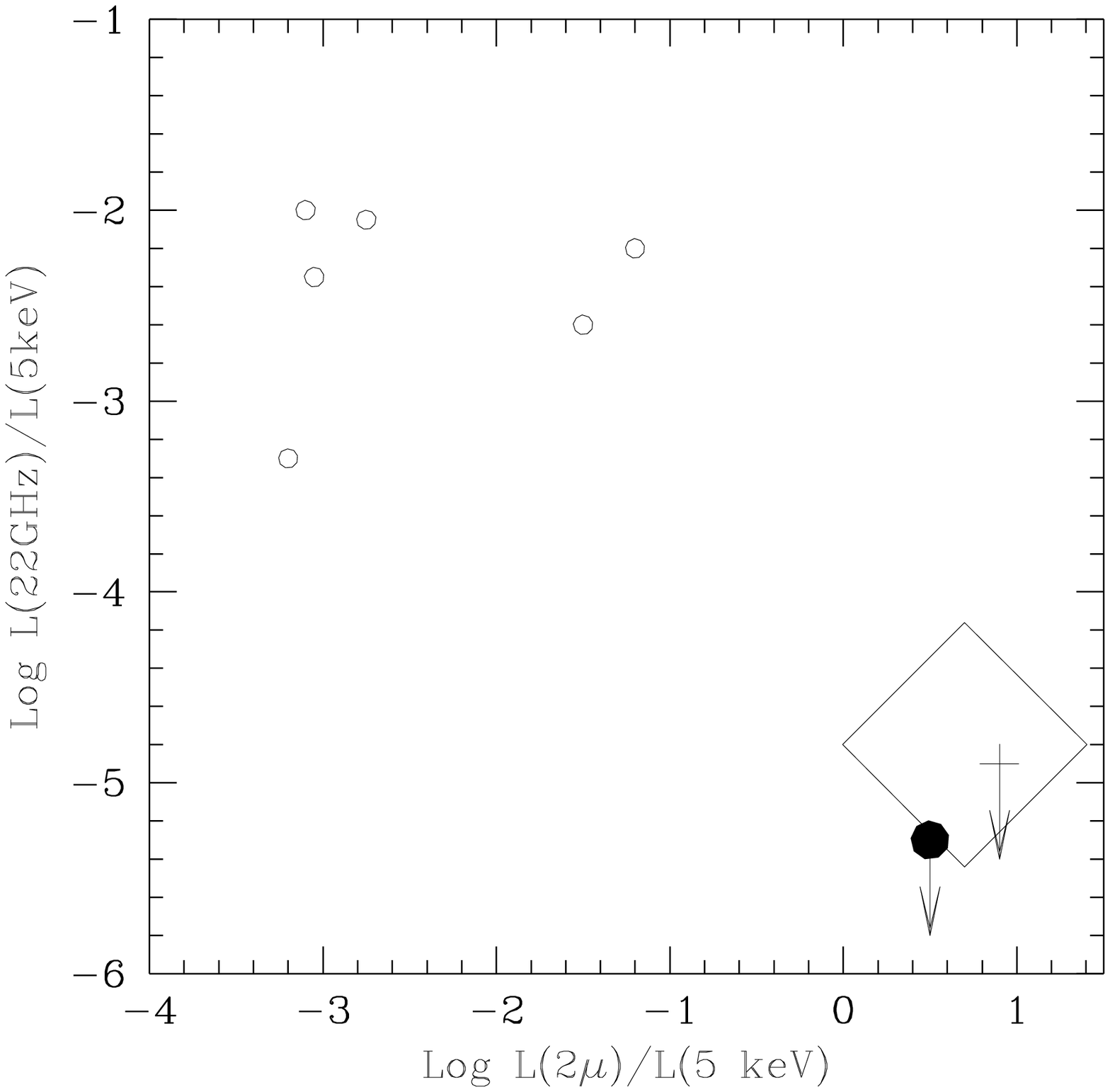,height=12cm,width=12cm,angle=0}
}
\caption{ The ratio between the 22GHz luminosity to the 5 keV
luminosity as a function of the ratio between the 2 micron luminosity
and the 5 keV luminosity. Big diamond = area covered by the radio quiet 
quasar SEDs of Elvis et al. (1994);
filled circle = NGC4258, BeppoSAX
observation; cross = NGC4258, ASCA observations (average); 
open circles = best fit ADAF plus wind models to the six  elliptical
galaxies studied by Di Matteo et al. (2000). It should be noted that
while the X-ray and the radio luminosities of the six models are very
close to measured values, the NIR points come just from the models.
}
\label{rxirx}
\end{figure}

As said in the introduction the SED of NGC4258 was compared with broad
band ADAF models by Lasota et al. (1996) and Gammie et al. (1999).
The fit of detailed ADAF models to the radio to X-ray SED is beyond
the scope of this paper.  Here we limit ourselves to consider the two
broad band spectral ratios $R_{RX}$= L(22GHz)/L(5keV) and
$R_{NIRX}$=L($2\mu$)/L(5keV), plotted in figure \ref{rxirx}. The
NGC4258 BeppoSAX and ASCA data have been combined with the Herrnstein
et al. (1998) 22 GHz upper limit and with the Chary \& Becklin (1997)
$2\mu$ ``nuclear'' emission. The resulting ratios are compared with
the analogous ratios built from the Di Matteo et al. (2000) best fit
ADAF + wind models to the SEDs of six nearby elliptical galaxies; and
with the ratios computed from the average SEDs of UVX selected radio
quiet quasars (Elvis et al. 1994).  From figure \ref{rxirx} we note
that the NGC4258 points are in the region occupied by radio quiet AGN;
and that the $R_{RX}$ and $R_{NIRX}$ values are 2-4 orders of magnitude
lower and higher, respectively, than in ADAF plus wind best fit
models.

\section{Discussion and Conclusions}

\subsection{Origin of the nuclear emission}

The nuclear luminosity L(2-10 keV) of NGC4258 during the BeppoSAX
observation was about $10^{41}$ erg s$^{-1}$, similar to the NIR $2\mu$
luminosity estimated by Chary \& Becklin (1997). This is consistent
with an extrapolation of the X-ray power law down to a few microns. On the
high energy side, the BeppoSAX data are not able to provide a strong
constraint on the cut-off. However, BeppoSAX measured a cut-off energy
of 100-200 keV in half a dozen Seyfert galaxies (Matt 2000 
and references therein; see also Zdziarski et al. 2000, and
references therein, for GRO/OSSE results). It is therefore reasonable to 
assume that the nuclear power law component can be extrapolated 
up to $\sim200$ keV. If this is the case the 3 micron to 200
keV total nuclear luminosity is of the order of $10^{42}$ erg
s$^{-1}$.  The uncertainty on this number strongly depends on the
assumed NIR to X-ray spectral index (an uncertainty of 0.2
on this spectral index would imply a $50\%$ uncertainty on the total
luminosity). The nuclear NIR to X-ray luminosity is similar to the
minimum {\it bolometric} luminosity implied by the Wilkes et
al. (1995) detection of polarized flux.  A nuclear NIR to X-ray 
luminosity of $10^{42}$ erg s$^{-1}$ and a black hole mass of
$3.9(\pm0.1)\times10^7$ M$_{\odot}$ (Herrnstein et al. 1999) imply an
Eddington ratio of $\approx0.0002$.  Although higher than the
Eddington ratios estimated by Lasota et al. (1996) and Gammie et
al. (1999) this is still in the ADAF regime (e.g. Narayan, Mahadevan \&
Quataert 1998). We therefore discuss briefly in the following the
applicability of ADAF models to the nuclear emission of NGC4258.

The SED of ADAF models depends on a variety of parameters and
assumptions (Narayan, Mahadevan \& Quataert 1998, Quataert \& Narayan
1999, Di Matteo et al. 2000).  Here we distinguish between two main
families of models, (a) those in which strong winds efficiently deplete
the emission region of emitting electrons and reduce the synchrotron
radio emission together with the inverse Compton emission in X-rays
(e.g. Di Matteo et al 2000), and (b) those without winds.  The observed
2.5-70 keV spectral index ($\alpha_E=1.11\pm0.14$) is steeper than
what expected in ADAF models where the contribution of Comptonization
is negligible and the X-ray luminosity is mostly due to thermal
bremsstrahlung with kT=100-200 keV, i.e. $\alpha_E\approx0.3$ at
energies below 50-100 keV (case (a), see e.g. di Matteo et al. 2000).
An additional kT$\approx100$ keV bremsstrahlung component can
contribute for at most one third of the broad band X-ray luminosity of
the main ($\alpha_E=1.11$) power law component (see Sect. 4.1).

On the other hand, if the X-ray band is dominated by Comptonization,
case (b), the observed X-ray spectral index can be easily
explained. In this case we would expect either strong synchrotron
emission in the radio band, or strong infrared emission from the outer
thin disk, to provide enough soft photons for the inverse Compton
scattering. We discuss these two possibilities in turn.

\noindent
{\bf Strong synchrotron emission}

Given the X-ray luminosity observed by ASCA and the higher one
observed by BeppoSAX, the Herrnstein et al. (1998) 22GHz upper limit
is barely consistent with the Gammie et al. (1999) ADAF models without
wind.  Furthermore, the observed radio-to-X-ray luminosity ratio is
2--3 orders of magnitude lower than in nearby elliptical galaxies
supposed to host an ADAF (see Figure \ref{rxirx} and Sect. 4.4).  ADAF
solutions with or without winds seem more ``radio-loud'' than the
NGC4258 nucleus.  Conversely, the radio-to-X-ray luminosity ratio is
consistent with that found in radio-quiet AGNs (Elvis et al. 1994). We
also note that the 22GHz upper limit is consistent with the luminosity
of a radio jet, whose power is proportional to the optical luminosity
of a standard accretion disc (as inferred from the optical emission
lines, Falcke \& Biermann 1999).

\noindent
{\bf Strong Infrared emission}

The NIR to X-ray ratio of NGC4258 is also similar to that of
radio-quiet AGNs (see figure \ref{rxirx}), while it is higher than in
the Di Matteo et al. (2000) ADAF models with winds by 2-4 orders of
magnitude.  In fact, Gammie et al. (1999) associate the strong nuclear
NIR luminosity detected by Chary \& Becklin (1997) to emission from a
thin disc extending down to 5-50 Schwarzschild radii ($R_S=2GM/c^2$).
According to Gammie et al. (1999) the flow should become advection
dominated and produce a geometrically thick configuration at this
transition radius.  The size of the X-ray emission region and
therefore the putative transition radius can be constrained through
X-ray variability studies. There are at least two relevant timescales
here, the local dynamical timescale $t_d=R^{3/2}(GM)^{-1/2}$ and the
causality timescale $t_c=R/c$. Large variations on timescales of
$\sim40,000$ seconds are present in the 3-10 keV light curve of
NGC4258, as well as 10-20\% variations on shorther ($\sim 1$ hour)
timescales.  Associating the longer timescales to $t_d$, as expected
if the hard X-ray luminosity is mostly due to bremsstrahlung emission,
produces an emission radius in units of $R_S$ of $r\ls17$.
Associating the same timescale to $t_c$ produces $r\ls100$. These
radii are still consistent with the ADAF transition radius of
$\sim5-50 R_S$ suggested by Gammie et al. (1999). We note however,
that even if a transition radius as small as $5 R_S$ may be
mathematically correct, it is less than two times the last stable
orbit in a Schwarzschild metric. Since the transition between a
geometrically thin disk to a thick configuration is unlikely to be
very sharp, this leaves little space for the putative hot plasma
region responsible for the X-ray emission.  Reynolds et al. (2000)
suggest that the size of the nuclear X-ray source may be as large as
$\sim50 R_S$, based on the limit on the width of the iron K$\alpha$
line. However, a sizeable fraction of the line emission may well be
produced by outer gas, as for example the gas responsible for the
X-ray absorption, see Sect. 4.1.

In conclusion, ADAF models with strong winds can be ruled out for the
nuclear emission of NGC4258, based on both the measured X-ray spectral
shape and the X-ray variability.  ADAF models without strong winds may
be applicable but it is not clear if they are a viable physical
solution (Blandford \& Begelman 1999, Di Matteo et al. 2000).
Moreover, they imply a radio to X-ray ratio barely consistent with the
measured upper limit, and also for them the X-ray observed variability
constrains the transition radius to be rather small (between 
$\sim20$ and 100 $R_S$).

The IR-to-X-ray Eddington ratio, the X-ray variability and spectral
shape, and the radio-to-X-ray and NIR-to-X-ray luminosity ratios
suggest that the nucleus of NGC4258 is an AGN in a low state (see
e.g. Siemiginowska et al. 1996) or a scaled-down version of a Seyfert
nucleus. This is also consistent with the conclusions of Neufeld \&
Maloney (1995) who suggest that the low luminosity of NGC4258 is the
result of low accretion rate ($\sim 10^{-4} M_{\odot} yr^{-1}$). In
the model of Neufeld \& Maloney (1995) the radiative efficieny is
high, $\sim 10 \%$, in contrast to the low efficiency ADAF models.
The X-ray nuclear luminosity can be naturally explained in terms of
Comptonization of soft photons in a hot corona (e.g. Haardt
\& Maraschi 1991, Haardt, Maraschi \& Ghisellini 1994).  Witt et
al. (1997) developed a model of accretion disc plus corona introducing
a coupling between the energy dissipated in the corona and the
accretion rate.  They found that if the viscosity mechanisms are the
same in the disc and in the corona, the corona does not form for
accretion rates smaller than a critical value, because Compton cooling
wins against viscous heating in the corona.  This critical value is
$\dot m_c\approx0.001$ at small radii and increases with the radius.
Under this condition, if the Eddington ratio of 0.0002 is linearly
proportional to $\dot m$ and if the efficiency in the conversion of
accretion power into radiation is $\eta\sim0.1$, a corona should not
form in the nucleus of NGC4258. There are at least two ways out from
this problem. The first is that the efficiency $\eta$ is at least as
small as 0.01. This may be the case if a large part of the accretion
power is carried out by a different channel, for example kinetic
energy in a wind (as also foreseen in the Witt et al. model).  Second,
the dissipation of accretion power and the viscosity mechanisms may be
very different in the disc and in the corona, as in the case of local
magnetic reconnection flares in the corona, envisaged by Haardt,
Maraschi \& Ghisellini (1994).

More theoretical work is clearly needed to understand if accretion
rates as small as 0.0002 in Eddington units may power a low luminosity
version of a normal Seyfert galaxy nucleus, although this already
seems for NGC4258 the most likely and less demanding, in terms of
complexity, scenario.

\subsection {Origin of the low energy component(s)}

Makishima et al. (1994) and Ptak et al (1999) found some support for a
two temperatures model for the low energy ASCA spectrum of this
source. Using ROSAT HRI data Pietsch et al. (1994) and Cecil, Wilson
\& De Pree (1995) were able to resolve the soft X-ray emission in
several components, the most relevant being a jet component.  Cecil,
Wilson and De Pree (1995) and Makishima et al. (1994) also report the
discovery of a hotter thermal component (kT$\sim4$ keV).  The BeppoSAX
data confirm the presence of at least two components, in addition to
the obscured nuclear power law, in the 0.1-10 keV spectrum of this
source.  One component must be thermal, given the strong iron L
complex detected, and its best fit temperature is kT$\sim0.6$ keV.  It
can be due to shocks formed in the interaction of the jet with the
interstellar matter (Cecil et al. 1995).

The shape of the second component is not well constrained by the
BeppoSAX data. It is consistent with both thermal emission with
temperature of kT$>1.3$ keV or a power law with energy index
$0.2^{+0.8}_{-0.2}$. This component may consist of a fraction of
0.935$\pm0.010$ of the nuclear power law spilling out from a non
uniform absorber.  NGC4258 is known to be highly polarized (Wilkes
etal 1995). This implies that there are lines of sight to the nucleus
scattering photons towards us and providing an unobscured view. Such
geometry mimicks a partial covering of the nuclear continuum.  The
observed covering fraction of is consistent with the 5--10 %
polarization observed in this source (Wilkes etal 1995).
Alternatively, the second component may be due to integrated emission
from binaries and Supernova Remnants. The contribution from these
discrete sources may be estimated from the B band luminosity of
NGC4258 (Trinchieri \& Fabbiano 1985, Canizares, Trinchieri \&
Fabbiano 1987, Fabbiano et al. 1992). The total B magnitude, corrected
for extinction, is B=8.53 (de Vaucouleurs et al 1991). This
corresponds to a luminosity in the B band, in units of the solar
luminosity, of log$L_B$=10.49.  The Fabbiano et al. (1992)
relationship between $L_X$ and $L_B$ for normal spiral galaxies
implies a 0.2-4 keV luminosity of $6\times10^{39}$ erg s$^{-1}$ from
discrete sources, which is $\sim 70\%$ the measured ``hot'' component
0.2-4 keV luminosity.  Since the scatter in the Fabbiano et al. (1992)
correlation is a factor of three we cannot exclude that the entire
observed ``hot'' component is due to discrete X-ray sources in the
host galaxy.  High spatial resolution observations, like those allowed
by the superior telescopes on board the Chandra satellite, will be
able to resolve the discrete sources in the galaxy and therefore
measure or put strong limits to any truly diffuse ``hot'' component.
 
\subsection{Strong X-ray absorption and water maser emission}

The nuclear continuum of NGC4258 varies on a timescale of half a day
(see Sect. 3). Our data are not good enough to detect Fe-K line
variability. However, all the observations to date are consistent with
Fe like flux being constant over at least few times 10$^5$ seconds. We
also find that the flux of the Fe line is consistent with its origin
in the absorbing gas (Sect. 4.1). All this places the Fe line, and so the
absorber at a distance of $>6\times 10^{15}$ cm. The water maser
emission in NGC4258 originates at 0.13 pc from the nucleus. Thus the
location of the X-ray absorber is consistent with being the same as
that of the maser emission.  Such an association between the X-ray
absorber and maser emission is further corroborated by a general
tendency of maser sources to show strong X-ray absorption, see for
example NGC4945 Circinus galaxy, NGC1068 (Matt. et al. 2000 and
references therein), ESO103-G35 (Wilkes et al. 2001) and possibly
NGC3079 (Bassani et al. 1999).

In the models of Neufeld \& Maloney (1995) for maser emission in
NGC4258, the X-rays from the nuclear AGN are shielded by gas of large
column density. Such shielding helps creating the layer of warm
molecular gas in the midplane of the circumnuclear disk. The column
density of the shielding gas in the Neufeld \& Maloney (1995) model is
$9\times10^{22}$ cm$^{-2}$, exacly what we measure in the present
BeppoSAX observation. It is thus reasonable to assume that the large
column density gas responsible for the nuclear X-ray absorption is
also shielding the molecular gas.  Neufeld \& Maloney (1995) also note
that if the circumnuclear disk is absolutely flat, i.e. not warped,
then it would not be illuminated by the X-ray continuum. In this
situation the temperature in the midplane of the disk would be too
cold to excite water maser emission. Note that in such a geometry there
would not be any X-ray absorption either.

All the above facts point towards a direct causal link between strong
X-ray absorption and water maser emission. Illumination by X-ray
continuum and subsequent shielding by a large column density gas seems
to be required for maser emission, at least in some models. If such a
scenario is correct, then every maser emitting AGN would show
absoption in its X-ray spectrum.

\bigskip

We would like to thank the BeppoSAX hardware teams for the development
and calibration of the instruments.  In particular, we would like to
remember the late Daniele Dal Fiume for his continuous dedication and
his fundamental contribution to an instrument of unprecedented
sensitivity and robustness like the BeppoSAX PDS.

This research has made use of SAXDAS linearized and cleaned event
files (Rev.2.0) produced at the BeppoSAX Science Data Center.  It has
been partially supported by ASI contract ARS--99--75 and MURST grant
Cofin--98--032.  SM work has been partly supported by the NASA grant
NAG5-8913 (LTSA).  We thank Tiziana Di Matteo, Fabrizio Nicastro and
Massimo Cappi for useful discussions.


\end{document}